\documentclass[pre,twocolumn,showpacs,superscriptaddress]{revtex4}
\usepackage{epsfig,amsmath,amssymb,graphics,color,calc}

\newcommand{\be}{\begin{equation}}
\newcommand{\ee}{\end{equation}}
\newcommand{\ba}{\begin{eqnarray}}
\newcommand{\ea}{\end{eqnarray}}
\newcommand{\ban}{\begin{eqnarray*}}
\newcommand{\ean}{\end{eqnarray*}}

\begin{document}

\title{Amorphous silica modeled with truncated and screened Coulomb 
interactions: A molecular dynamics simulation study}
\date{\today}

\author{Antoine Carr\'e}

\affiliation{Institut f\"ur Physik, Johannes--Gutenberg--Universit\"at Mainz,
Staudinger Weg 7, 55099 Mainz, Germany}

\affiliation{Laboratoire des Collo{\"\i}des, Verres et Nanomat{\'e}riaux,
UMR 5587, Universit{\'e} Montpellier II and CNRS, 34095 Montpellier, France}

\author{Ludovic Berthier}

\affiliation{Laboratoire des Collo{\"\i}des, Verres et Nanomat{\'e}riaux,
UMR 5587, Universit{\'e} Montpellier II and CNRS, 34095 Montpellier, France}

\affiliation{Joint Theory Institute, Argonne National Laboratory and
University of Chicago,
5640 S. Ellis Av., Chicago, Il 60637}

\author{J\"urgen Horbach}

\affiliation{Institut f\"ur Physik, Johannes--Gutenberg--Universit\"at Mainz,
Staudinger Weg 7, 55099 Mainz, Germany}

\affiliation{Institut f\"ur Materialphysik im Weltraum, Deutsches Zentrum f\"ur
Luft-- und Raumfahrt, 51147 K\"oln, Germany}

\author{Simona Ispas}

\affiliation{Laboratoire des Collo{\"\i}des, Verres et Nanomat{\'e}riaux,
UMR 5587, Universit{\'e} Montpellier II and CNRS, 34095 Montpellier, France}

\author{Walter Kob}

\affiliation{Laboratoire des Collo{\"\i}des, Verres et Nanomat{\'e}riaux,
UMR 5587, Universit{\'e} Montpellier II and CNRS, 34095 Montpellier, France}

\begin{abstract}
We show that finite-range alternatives to the standard 
long-range BKS pair potential for silica might be used 
in molecular dynamics simulations. We study two 
such models that can be efficiently simulated since
no Ewald summation is required.
We first consider the Wolf method, where the Coulomb interactions
are truncated at a cutoff distance $r_c$ such that the requirement of
charge neutrality holds. Various static and dynamic quantities are
computed and compared to
results from simulations using Ewald summations.
We find very good agreement for $r_c\approx 10$\,\AA.  
For lower values of $r_c$, the long--range
structure is affected which is accompanied by a slight acceleration of
dynamic properties. 
In a second approach, the Coulomb interaction is
replaced by an effective Yukawa interaction with two new
parameters determined by a force fitting procedure. 
The same trend as for the Wolf method is seen. However, 
slightly larger cutoffs have
to be used in order to obtain the same accuracy with respect to static
and dynamic quantities as for the Wolf method.
\end{abstract}


\maketitle

\section{Introduction}

Silica (SiO$_2$) is the prototype of a glassformer that exhibits a
tetrahedral network structure. It is the basic oxidic component of many
minerals and technological glasses. In recent years, molecular
dynamics (MD) computer simulations have provided valuable insight
into static and dynamic properties of amorphous and crystalline silica
\cite{poole95,vollmayr96,horbach99_1,horbach99_2,horbach01,horbach02,
scheidler01,roder01,benoit01,benoit02,berthier,taraskin97,muser01,
herzbach05,saika01,saika04_1,saika04_2,shell02,saksa06,liang07,ma06,
kerrache06,leonforte06,berthier07,berthier07b}.
In a  classical MD simulation, the interactions between the atoms are described
by an effective  potential where, different from {\it ab initio}
approaches, the electronic degrees of freedom are not taken into account
explicitly. In many of the aforementioned MD studies, the so--called BKS
potential \cite{beest90} has been used. Although it is a standard pair
potential, it yields good agreement with experimental
data. However, it contains a long--range Coulomb interaction term and
thus the computation of energies and forces is very expensive, in 
particular for large systems. 
Therefore, it is important to know whether silica
can also be modeled by a potential with a {\it finite} range.

The classical approach to evaluate Coulomb energies and forces in
a simulation is the Ewald summation method where the sum over all
Coulomb interactions is decomposed into a real space and a Fourier
part \cite{ewald21,allen87,frenkel02}. For a three--dimensional
$N$ particle system with periodic boundary conditions in all three
spatial directions, Ewald method yields a $N^{3/2}$ scaling for the
computational load \cite{perram88,frenkel02}. Even worse is the case of
a quasi--two--dimensional slab geometry for which Ewald method exhibits
a $N^2$ scaling \cite{parry75}.  There are modifications of the Ewald
method such as particle--particle particle--mesh methods (PPPM) 
\cite{hockney88,deserno98}
that yield a scaling proportional to $N{\rm log}N$, but these methods
cannot be implemented as efficiently as in the three--dimensional case for
the quasi--two--dimensional geometry \cite{arnold02}. A scaling better
than ${\cal O}(N{\rm log}N)$ can be reached by means of multipole methods
\cite{greengard87}. However, these methods introduce a large computational
overhead such that their use is only reasonable for a very large number
of particles, say $N>10^5$~\cite{esselink95}.

Apart from the case of quasi--two--dimensional geometries, there are
other applications where it is difficult to handle long--range Coulomb
interactions. For instance, the calculation of transport coefficients
such as the shear viscosity via Green--Kubo relations is in general more
complicated and less efficient when Ewald sums have to be considered
\cite{allen87}. This is also the case for Monte Carlo (MC) simulations.
When using Ewald summation (or also PPPM) in MC simulations, the full
Fourier part of the Coulomb energy must be evaluated after each particle
displacement by summing over $N$ terms and 
$N_k$ different wavevectors used in the Ewald summation.
Note that this is much worse than for MD where the sum over wavevectors
can at least be used to compute the force over all $N$ particles, while
the energy must be recomputed after each single move in MC 
simulations.

For all these reasons, a reliable finite range pair potential 
to simulate silica is highly desirable.
In this paper, we address this issue by making the 
assumption that, due to screening effects,
the long--range Coulomb interactions in typical ionic systems can in fact be
truncated~\cite{wolf92}. 
However, a cutoff distance $r_{\rm c}$ for a $r^{-1}$ potential
cannot be introduced in a similar manner as for a short--range potential
since a truncated spherical summation over pairwise $r^{-1}$ Coulomb
terms leads to a violation of charge neutrality. In order to circumvent
this problem, Wolf {\it et al.}~\cite{wolf92,wolf99} have introduced
a simple correction term in the truncated Coulomb sums that recovers
the requirement of charge neutrality. 
The Wolf method~\cite{wolf92,wolf99,fennell06} 
has been used in recent simulation studies, e.g.
for quasi--two--dimensional geometries \cite{gallo02}, in MC simulations
\cite{avendano06}, or for dipolar fluids \cite{moreno06}.  
Of course, it is {\it a priori}
not clear how large the cutoff radius $r_{\rm c}$ has to be in order
to correctly reproduce the static and dynamic properties of the original
model with long--range Coulomb interactions.  We have 
carefully studied this issue using  
MD simulations of amorphous silica based on the
BKS potential. We show that a cutoff of about 10\,\AA~is
necessary to obtain good agreement with the initial long--range model on
a quantitative level. This indicates that, due to screening effects,
amorphous silica can indeed be described by an effective potential of
finite range. Therefore, in our second approach we reparametrize the
Coulomb term of the BKS potential by replacing it by a screened Coulomb
(Yukawa) potential with two new
parameters. The parameterization of this Yukawa potential is achieved
by a force fitting procedure based on previous MD simulations of BKS
silica \cite{horbach99_1}. This alternative procedure is 
of course more involved than the simpler Wolf truncation, as
the new pair potential must be carefully parameterized.

The paper is organized as follows. In Sec.~\ref{models} we introduce
the truncated and screened Coulomb potentials studied in this work,
and we describe how the free parameters are fixed. In Sec.~\ref{results}
we use a given set of fixed parameters and compare in detail static
and dynamic properties of the original long--range BKS model, and the
two finite--range alternatives suggested in this paper.  We present our
conclusions in Sec.~\ref{conclusion}.

\section{Two finite--range alternatives}
\label{models}
\subsection{The original long-range BKS model}

The functional form of the BKS potential is~\cite{beest90}
\begin{equation} 
\phi_{\alpha \beta}^{\rm BKS}(r)= 
{q_{\alpha} q_{\beta} e^2}V_C(r) + 
A_{\alpha \beta} \exp\left(-B_{\alpha \beta}r\right) - 
\frac{C_{\alpha \beta}}{r^6}, 
\label{bks_pot} 
\end{equation} 
where $\alpha, \beta \in [{\rm Si}, {\rm O}]$, $r$ is the distance
between the ions of type $\alpha$ and $\beta$. 
The values of the
constants $q_{\alpha}, q_{\beta}, A_{\alpha \beta}, B_{\alpha \beta}$,
and $C_{\alpha \beta}$ can be found in Ref.~\cite{beest90}. For the sake
of computational efficiency the short range part of the potential was
truncated and shifted at 5.5\,\AA~\cite{vollmayr96}. This truncation
also has the benefit of improving the agreement between simulation and
experiment with respect to the density of the amorphous glass at low
temperatures~\cite{vollmayr96,horbach99_1}. Finally, the function
\begin{equation} 
V_C(r) =\frac{1}{r}
\label{coulomb} 
\end{equation} 
is the Coulomb long-range term which will be further approximated by finite
range potentials. The other terms in Eq.~(\ref{bks_pot}) will be unchanged.

In numerical simulations using periodic boundary conditions, 
 the Coulombic part of the potential energy, $E_{\rm
coul}$, is given by the following formula:
\begin{equation}
  E_{\rm coul} = \frac{1}{2} \,
               \sum_{{\bf n} \in {\bf Z}^3 } \,
               \sum_{i,j=1 \atop i \neq j \; \; {\rm for} 
                     \; \; {\bf n}=0}^{N} \,              
              \frac{q_i q_j e^2}{|{\bf r}_{ij} + {\bf n}L|} \ .
  \label{coulpot}
\end {equation} 
Here, ${\bf r}_{ij} = {\bf r}_i - {\bf r}_j$ is the distance vector
between particle $i$ and particle $j$. The sum over ${\bf n}$ takes into
account the interactions of a particle $i$ with all the replicated image
particles (including the images of particle $i$).
Moreover, it is only conditionally convergent,
i.e.~the value of the sum depends on the order by which the terms are
summed up. An efficient method to circumvent these problems is provided
by the Ewald summation technique. However, finite range alternatives
to Ewald sums are very desirable, as outlined in the Introduction. The
analysis of such alternatives is the principal aim of this paper.

We shall analyze the ability of finite--range potentials to reproduce
the behavior of the original BKS model by comparing the results of
molecular dynamics simulations of the new potentials against results
employing Ewald sums.  The latter results are taken from previously
published work~\cite{horbach99_1,berthier}
 and they will be labelled ``Ewald simulations'' hereafter.  
 Most of the Ewald simulations
were performed with $N=1008$ particles, at the density 2.37\,g/cm$^3$,
with Ewald parameters ($\alpha$ and $N_k$) optimized as in 
\cite{vollmayr96}.
In particular the real space part of the Coulomb force is truncated
at $10.17$~\AA, and we take the results of these simulations as
representative of the behavior of the ``real'' Coulomb potential. We know,
however, that finite size effects are present for this system size,
and we borrow additional data from the $N=8016$ particles simulations
from \cite{horbach99_1} when necessary.

\subsection{Truncation using the Wolf method}

\label{wolfmethod}

\begin{figure}
\psfig{file=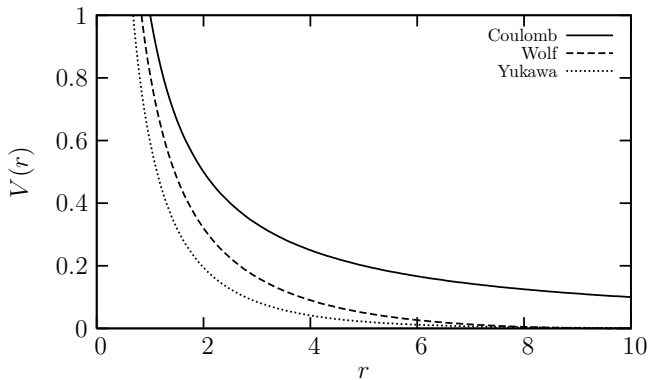,width=8.5cm}
\caption{\label{fig1} Comparison of the pure $1/r$ Coulomb interaction,
the Wolf truncation using Eq.~(\ref{wolf}) for $r_c=10.17$~\AA, and the Yukawa 
potential in Eq.~(\ref{eq_yuk}), for $D_{\rm Y}=1.07$, $\Delta=5.649$~\AA,
$r_c=10.17$~\AA.}
\end{figure}

As proposed in Refs.~\cite{wolf92,wolf99,fennell06}, the 
$V_C (r)$ term in the BKS potential (\ref{bks_pot}) may be approximated by the
following form:
\begin{equation}
V_W(r) = \left( \frac{1}{r} -\frac{1}{r_c} \right) 
+ \frac{1}{r_c^2} (r-r_c), \quad r<r_c, 
\label{wolf}
\end{equation}
while $V_W(r\ge r_c)=0$. The potential (\ref{wolf}) is a finite--range
potential: Only particles separated by distances smaller than $r_c$ 
interact. Such a potential becomes computationally extremely useful
when system sizes that are much larger than $r_c$ are studied. 
But Wolf's main point is that reasonable values of $r_c$ can lead to 
numerically accurate
results~\cite{wolf92}.  That this is by no means a trivial statement
can be appreciated in Fig.~\ref{fig1} which compares the initial
Coulomb interaction in (\ref{bks_pot}) to the expression given by Eq.\,(\ref{wolf}).
Both potentials are very different, and the truncation should in principle
have a drastic effect. Indeed previous naive truncation attempts have been
shown to produce quantitatively inaccurate results~\cite{kerrache06}.
We shall demonstrate, however, that the Wolf method performs well in
the case of liquid silica.

The truncated form of the potential (\ref{wolf}) can be justified as
follows. Wolf~\cite{wolf92} showed that the main error made when imposing
naively a finite distance cutoff to a $1/r$ potential stems from the
fact that a sphere of radius $r_c$ centered around a particle is in
general charged, so that its interaction with the rest of the system is
not negligible.  He proposed to approximate this interaction by using
the potential
\begin{eqnarray*} 
V(r) = \frac{1}{r} - \frac{1}{r_c}, \quad r<r_c,
\end{eqnarray*}
which amounts to screening completely  the charge  contained in the sphere
by placing the opposite charge on its surface.
This form is also natural from the computational point of view:
Truncated potentials in molecular dynamics simulations are always
shifted to avoid energy discontinuities at the cutoff, $V(r=r_c)=0$.
Although the shift is sufficient to compute the energy, which was the
initial problem considered by Wolf~\cite{wolf92}, this potential is not
well--suited for MD simulations since the forces are discontinuous at
$r_c$~\cite{wolf99}. Hence, the second term is added in Eq.~(\ref{wolf}),
as suggested in Ref.~\cite{fennell06}.

%More elaborate alternatives to the truncation in (\ref{wolf}) were also
%suggested, which employ additional screening functions and modify the
%original Coulomb interaction more profoundly. These methods are not
%well justified, and the detailed studies in \cite{fennell06} suggest that
%they do not yield qualitative improvements over the Wolf ansatz (\ref{wolf}).

We have performed MD simulations with the BKS model (\ref{bks_pot})
where we have approximated the $1/r$ Coulomb interaction term by the Wolf
formula, Eq.~(\ref{wolf}).  For the cutoff $r_c$ in (\ref{wolf}), three
different values have been chosen, namely $r_c=6.0$\,\AA, 8.0\,\AA,
and 10.17\,\AA. Note that a cutoff of $r_c =10.17$\,\AA~has been also
used for the real space part of the Ewald sums in our simulations 
using the original BKS potential \cite{berthier}. 
For a system of $N=1008$
particles, the gain in CPU time with the Wolf method is a factor of 2
for $r_c =10.17$\,\AA. Although this speed--up is not that impressive, one
should keep in mind that the Wolf method is also well--suited for problems
where Ewald sums become very inefficient, e.g.~for quasi--two--dimensional
geometries, for MC simulations, or for large systems (see Introduction).
In addition the structure of the code 
and the calculations of physical quantities
become much simpler if the potential has only
a finite range.

\begin{figure}
\psfig{file=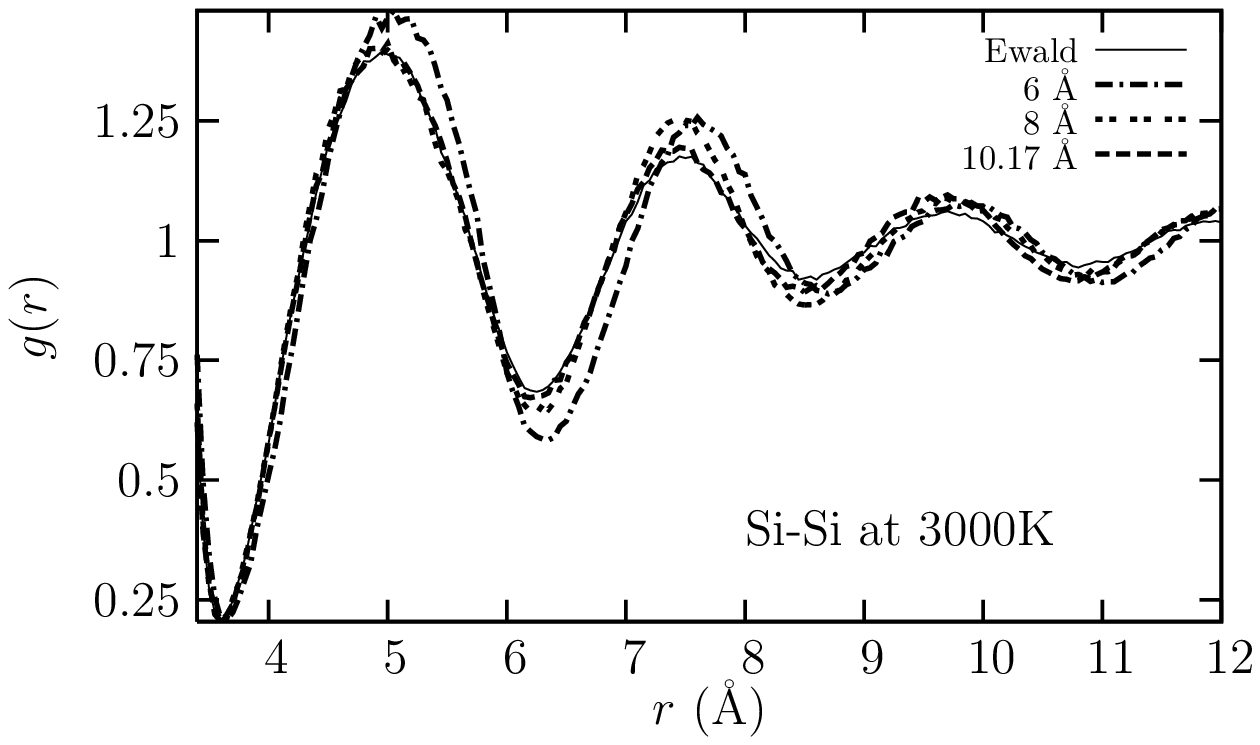,width=8.5cm}
\psfig{file=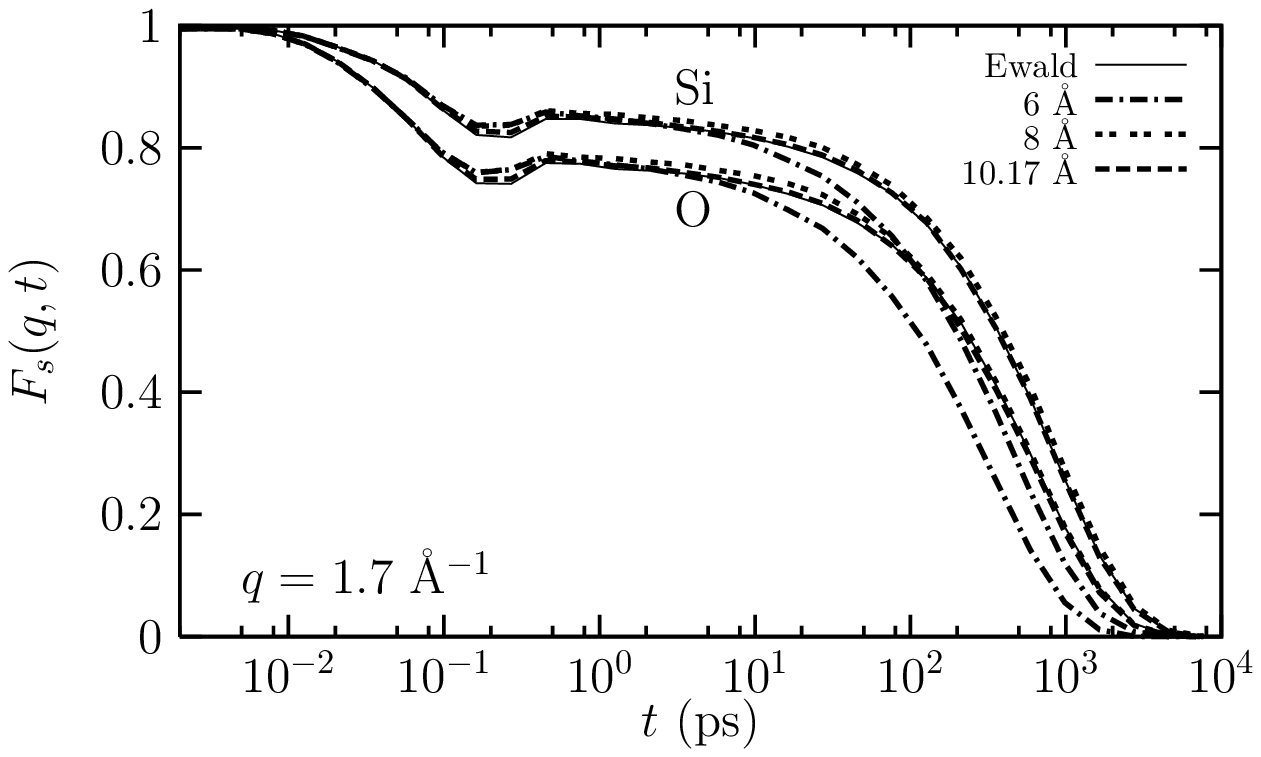,width=8.5cm}
\caption{\label{fig2} Top: Influence of the Wolf cutoff 
on the Si-Si pair correlation function for large distance.
Bottom: Influence of the Wolf cutoff on the 
time dependence of the self-intermediate scattering function
for both Si (top curves) and O (bottom curves). 
Both panels are for $T=3000$~K.}
\end{figure}

Equilibrated configurations from simulations with Ewald sums
\cite{berthier} were used as starting configurations for the MD
simulations with the Wolf method at each temperature between 3000\,K and
6100~K. 
The considered temperatures were 3000\,K, 3100\,K, 3250\,K, 3400\,K, 3580\,K, 
3760\,K, 4000\,K, 4300\,K, 4700\,K, 5200\,K and 6100\,K, i.e. the same values as
the ones considered in Ref.~\cite{horbach99_1}.
 As in Refs.~\cite{horbach99_1,berthier}, the velocity form of
the Verlet algorithm \cite{allen87} was used to integrate the equations
of motion with a time step of 1.6\,fs.  Equilibration runs were made
by coupling the system to a stochastic heat bath, before performing
productions runs in the microcanonical ensemble. During the equilibration,
we detect no systematic drift of the energy for $r_c=10.17$~\AA, which
is a first indication that Ewald and Wolf method should yield very close
results for this cutoff value.  During microcanonical simulations at 3000\,K, the
drift in the total energy is comparable in amplitude to the one observed
when using Ewald summations. Therefore we follow Ref.~\cite{horbach99_1} and
rescale velocities every $10^6$ time steps to maintain the total
energy constant.

We have analyzed a number of static and dynamic quantities (see
Sec.~\ref{results}), and we find that deviations between Wolf and Ewald
methods become more pronounced when temperature decreases, and, of course,
when the cutoff value decreases.
This temperature evolution suggests that charges are 
more efficiently screened at high temperatures 
in amorphous silica when the system is more disordered.

In Fig.~\ref{fig2} we present selected static and dynamic results
at $T=3000$~K for different values of the cutoff.  As a dynamic
quantity, we show the self part of the intermediate scattering function
\cite{glassbook},
\be
F_s({\bf q},t) = \frac{1}{N_\alpha} \sum_{j=1}^{N_\alpha} 
\left\langle {\rm e}^{i {\bf q} \cdot ({\bf r}_j(t) - {\bf r}_j(0))}
\right\rangle,
\ee
where ${\bf r}_j(t)$ is the position of particle $j$ of type $\alpha$ at time
$t$. As a representative structural quantity we show the pair correlation
function $g_{\alpha \beta}(r)$ (with $\alpha, \beta = {\rm Si, O}$),
\be
g_{\alpha \beta}(r) = \frac{V}{N_{\alpha} (N_{\beta}-\delta_{\alpha\beta})} 
\left\langle  \sum_{i=1}^{N_\alpha} \sum_{j=1}^{N_\beta}
  \frac{1}{4\pi r^2}
\delta (r - |{\bf r}_i - {\bf r}_j|) \right\rangle
\ee
where $V$ is the total volume of the system and $N_{\alpha}$ the number
of particles of type $\alpha$. In Fig.~\ref{fig2}, the function $g_{\rm
SiSi}(r)$ is shown  for $ r \ge 4 \mathrm \AA$ since the Si--Si correlations exhibit the largest
effects with respect to the cutoff $r_c$, at large distance.

For $r_c=6$~\AA, deviations between Ewald and Wolf methods are obvious.
Self-intermediate scattering functions decay faster with Wolf than 
with Ewald method, and the
positions of the second and third peaks in the pair correlation function
$g_{\rm SiSi}(r)$ arise at different values in the two methods.  
For $r_c=8$~\AA, the situation is already
much better since structural relaxation occurs at the correct timescale,
and oscillations in $g_{\rm SiSi}(r)$ are in phase with the corresponding
Ewald result. However, a closer inspection of the data shows that 
the plateau value in the self--intermediate
scattering function is not correctly reproduced and the amplitude of the
oscillations in $g_{\rm SiSi}(r)$ is too large at large distance.
For $r_c=10.17$~\AA,
the agreement with the Ewald simulation is almost perfect both for statics
and dynamics. The height of the plateau in $F_s(q,t)$ is now correct and
the amplitude of the oscillations in $g_{\rm SiSi}(r)$ are very close to
those of the Ewald result.  Looking very closely at the data, however,
a few deviations remain. The oscillations in $g_{\rm SiSi}(r)$ at large
distances are still too pronounced, meaning that long--range order in
the liquid is slightly more pronounced when using Wolf's method (see also the
discussion concerning the results shown in Fig. \ref{gr2}).
The amplitude of the small dip in $F_s(q,t)$ around $t~\sim 1$\,ps is much
smaller for $r_c=6$\,\AA~and 8\,\AA, and remains a bit too small for
$r_c=10.17$~\AA, although it is only a tiny difference in the latter case.

The influence of the cutoff and convergence towards the Ewald results
is further confirmed by the behavior of the pressure, which will be
studied in more detail in Sec.~\ref{results}.  In fact, we find that the
pressure represents a very sensitive test for the choice of $r_c$.
For $T=3000$~K, we find $P \approx -3.2$\,GPa, $-0.11$\,GPa, and
$0.83$\,GPa for $r_c=6$\,\AA, 8\,\AA, and 10.17~\AA, respectively, while
$P \approx 0.89$\,GPa using Ewald sums for the same number of particles.

From the results described in this section we decide therefore that
$r_c=10.17$\,\AA~is a good compromise between a small cutoff which
improves computational efficiency and a very large cutoff which
matches best the behavior observed in simulations using Ewald sums.
In Sec.~\ref{results}, we shall present a more extensive set of static and
dynamic data for $r_c=10.17$~\AA~and we will show that this produces a
physical behavior in satisfying quantitative agreement with simulations
using Ewald sums.

\subsection{Yukawa screening}
The replacement of the $1/r$ interaction by a screened
Coulomb interaction has been proposed in several recent
simulation studies of various atomistic systems (see, e.g.,
Refs.~\cite{ma06,kerrache06,zahn02,fennell06} and references therein). In
these studies, screened Coulomb or Yukawa potentials have been considered
as alternatives to the Wolf method presented above.

In the present work we reparametrize the BKS potential by replacing $V_C (r)$
term in Eq.~(\ref{bks_pot}) by a Yukawa interaction term of the following form,
\begin{equation}
  V_{\rm Y}(r)=D_{\rm Y} \frac{\exp\left(- r / \Delta \right)}{r}
                  , %q_{\alpha}q_{\beta}e^2
\label{eq_yuk}
\end{equation}
which introduces the amplitude $D_{\rm Y}$ and the screening length
$\Delta$ as new parameters. 
Note that in Eq.~(\ref{eq_yuk}) the same parameters 
$D_{\rm Y}$ and $\Delta$ are used for Si--Si, Si--O, and
O--O interactions. The shape of the Yukawa interaction is compared to the one  
of the Wolf and Coulomb terms in Fig.~\ref{fig1}.

\begin{figure}
\begin{center}
\psfig{file=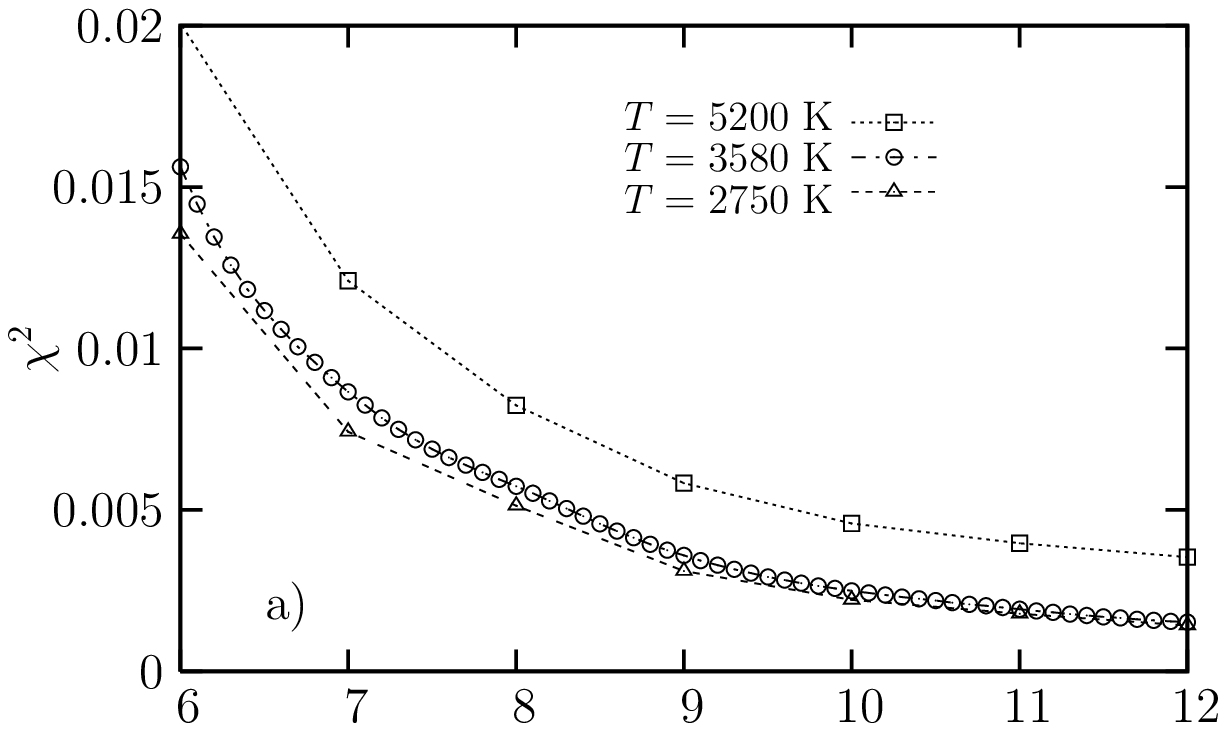,width=8.5cm}
\psfig{file=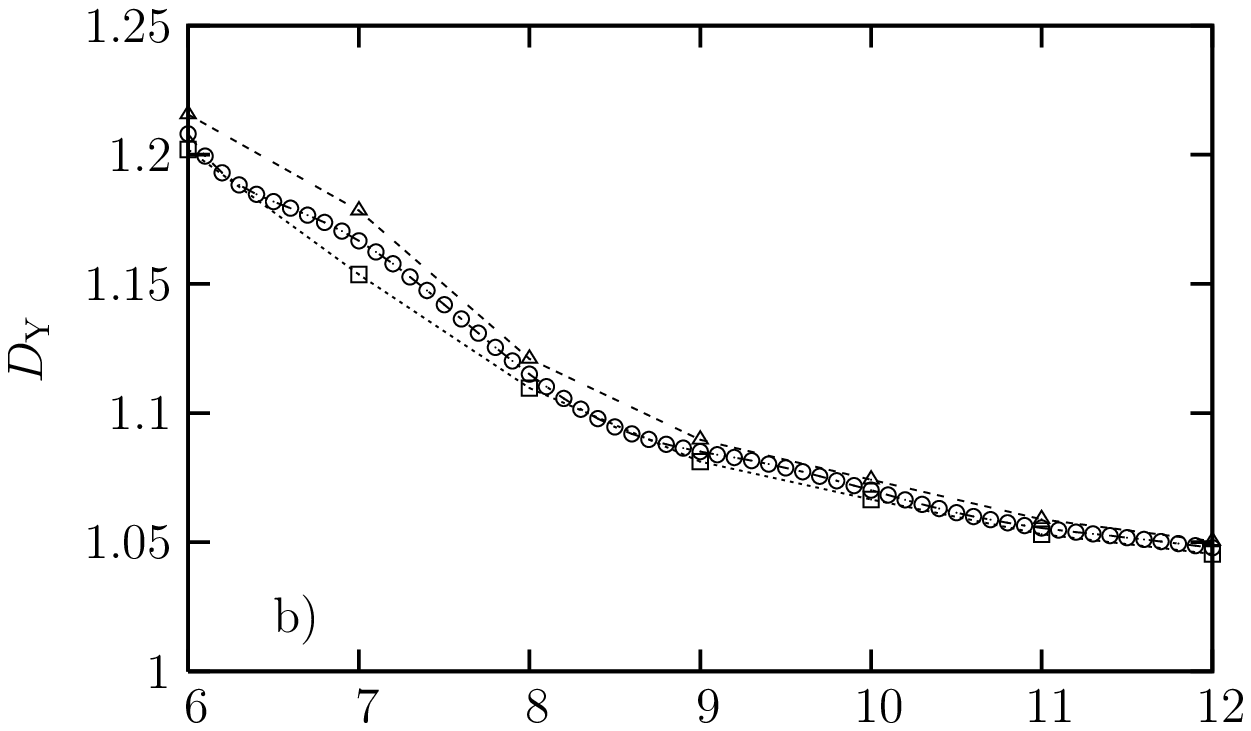,width=8.5cm}
\psfig{file=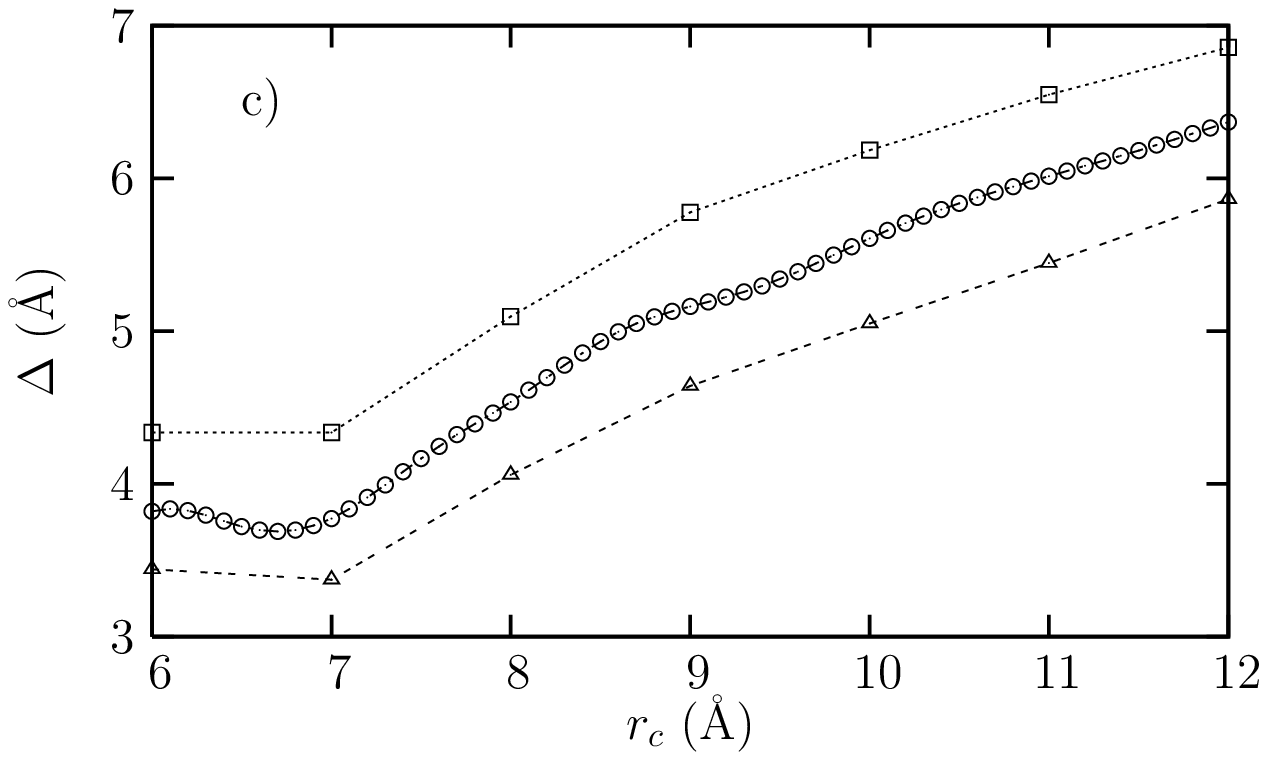,width=8.5cm}
\end{center}
\caption{\label{fig3} Dependence of the $\chi^2$, (a), 
prefactor $D_{\rm Y}$, (b), and screening length $\Delta$, (c), 
of the Yukawa  potential (\ref{eq_yuk}) as functions of the
cutoff $r_c$ for the three indicated temperatures.}
\end{figure}

Our determination of $D_{\rm Y}$ and $\Delta$ was based on previous
MD simulations for a system of 8016 particles using the BKS potential
\cite{horbach99_1}. At each temperature in the interval 6100\,K\,$\le T
\le$\,2750\,K (see Ref.~\cite{horbach99_1} for the considered values)
the parameters $D_{\rm Y}$ and $\Delta$ are fitted such
that the forces on each particle of the BKS configurations are optimally
reproduced by the new potential. More precisely, the fitting 
procedure was based
on the following $\chi^2$ function,
\begin{equation}
 \label{eq_chisq}
  \chi^2 = \left\langle \frac{1}{3N}  \sum_{\alpha={\rm Si},\,{\rm O}} 
\frac{1}{\sigma_\alpha^2}
 \sum_{i=1}^{N_\alpha}
          | {\bf F}_i^{\rm BKS} - {\bf F}_i^{\rm Y} |^2%}
             %  {\langle ({\bf F}_i^{\rm BKS})^2 \rangle - 
              %  \langle {\bf F}_i^{\rm BKS} \rangle^2}
	       \right\rangle,
\end{equation}
where ${\bf F}_i^{\rm BKS}$ and ${\bf F}_i^{\rm Y}$ denote the total
forces on the particle $i$ for the original BKS potential and the new
BKS potential modified by the screened Coulomb term (\ref{eq_yuk}),
respectively. The $\sigma_\alpha$ 
represents the standard deviation of the BKS force
distribution for Si and O particles. 
We find 
$\sigma_{\rm Si}(T=6100~{\rm K})=4.427614$ eV/\AA,
$\sigma_{\rm O}(T=6100~{\rm K})=3.674404$ eV/\AA,
$\sigma_{\rm Si}(T=2750~{\rm K})=3.235584$ eV/\AA,
$\sigma_{\rm O}(T=2750~{\rm K})=2.632564$ eV/\AA.
The brackets $\langle \cdots \rangle$ denote an average over
different samples (i.e.~BKS configurations).  

It is important to note that if we use the ansatz in 
Eq.~(\ref{eq_yuk}) with 
$\Delta$ and $D_{\rm Y}$ as the only two free parameters, the function $\chi^2$
could only be minimized by approaching the limits 
$\Delta=\infty$ and $D_{\rm Y}=1$,
thus reproducing the original Coulomb interaction.
Therefore, we have to introduce a third parameter, namely a 
cutoff distance $r_c$ above which the
potential $V_{\rm Y}(r)$ is set to zero. 
Thus, we shall first fix the value of the cutoff $r_c$, and then
determining the best set of parameters for 
$D_{\rm Y}$ and $\Delta$ to minimize the 
function in Eq.~(\ref{eq_chisq}). This procedure can then be repeated 
for different values of the cutoff $r_c$, and for different 
temperatures. The outcome of this study will be the suggestion  
of an ``optimal'' 
a set of parameters $(r_c, D_{\rm Y}, \Delta)$ 
that can be used to study the properties of silica
at various temperatures.  
 
Note that the use of a cutoff in the potential 
(\ref{eq_yuk}) implies a discontinuity of the forces at $r_c$.
As already mentioned in the previous section, this is not useful for MD
simulations since it leads to an energy drift in microcanonical simulation
runs. Therefore, we have multiplied the {\it forces} ${\bf F}_i^{\rm Y}$
by the function
\begin{equation}
  f(r) = \exp\left( - \frac{h^2}{(r-r_c)^2} \right),  \quad \quad r< r_c, 
\end{equation}
which makes the forces continuous at $r_c$. The constant $h$ was set
to $2.0$\,\AA.

The $\chi^2$ function (\ref{eq_chisq}) was minimized by means of the
Levenberg--Marquardt algorithm \cite{recipes} which is essentially a
conjugate gradient scheme for nonlinear fitting problems.  As shown
in Fig.~\ref{fig3}a, $\chi^2$ decreases when temperature decreases, although
the temperature dependence is relatively weak. It also decreases 
relatively quickly as
a function of the cutoff radius $r_c$ for the Yukawa potential. Empirically 
we find that the data at $T=3580$~K in Fig.~\ref{fig3}a can be well described
with a power law behavior, $\chi^2(r) \propto 1/r^{3.4}$.
It is indeed expected that $\chi^2$ vanishes at large $r_c$ 
since in the limit $r_c \to \infty$, the original Coulomb potential should
be recovered. Therefore, the
screening length $\Delta$ should diverge for large values
of $r_{\rm c}$ while the amplitude $D_{\rm Y}$ should approach 1, 
as confirmed by our numerical results in Figs.~\ref{fig3}b and c.

Whereas the temperature dependence of $D_{\rm Y}$ is relatively weak, 
the screening length $\Delta$ 
decreases significantly with decreasing temperature, which can be understood
as follows.  Physically, the value of
$\Delta$ obtained using the minimization procedure results from 
a compromise. On the one hand, a 
large screening length $\Delta$ should be used   
to recover the ``bare'' Coulomb potential. On the other hand, 
the introduction of a finite distance cutoff $r_c$ produces 
large errors due to incorrect charge balance~\cite{wolf92}.   
In the present scheme, this is compensated by screening more strongly 
the Coulomb interaction by using a smaller screening length
$\Delta$. 
Thus the results in Fig.~\ref{fig3}c indicate that charge neutrality 
is best satisfied, for a fixed cutoff value $r_c$, when 
temperature is higher, so that a larger screening length
$\Delta$ can be used at high temperature. In turn, this suggests
that charges in amorphous silica are screened 
on a smaller lengthscale when temperature is high, that is, 
when the system is more disordered. The same conclusion 
was reached in Sec.~\ref{wolfmethod}. 

\begin{figure}
\psfig{file=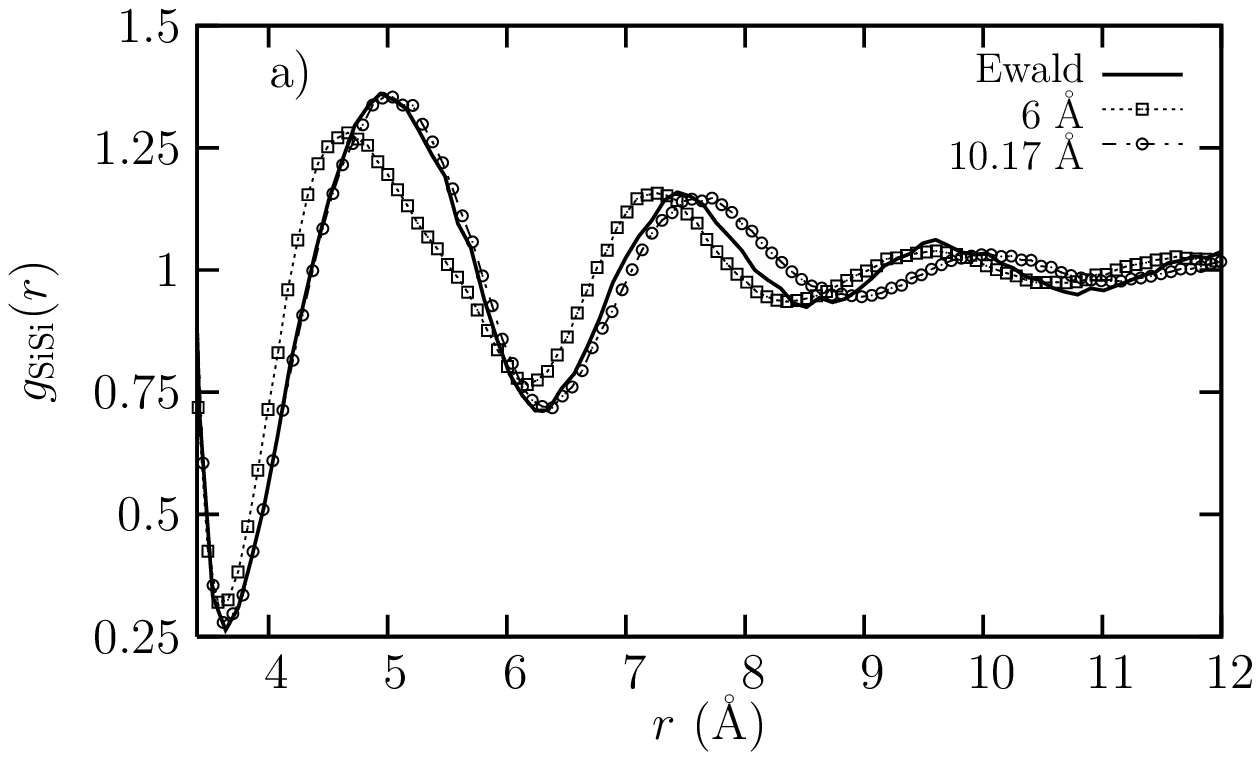,width=8.5cm}
\psfig{file=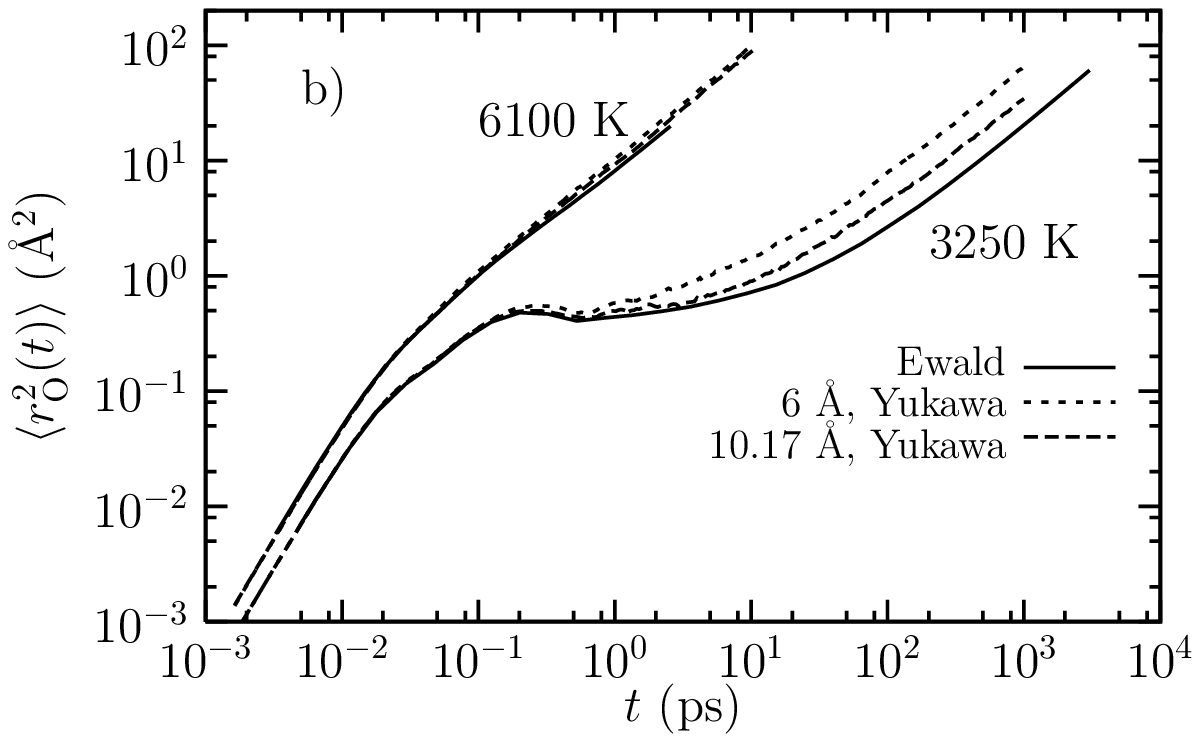,width=8.5cm}
\caption{\label{fig6} Influence of the cutoff $r_c$ for 
the Yukawa potential 
on the Si-Si pair correlation function at $T=3250$~K (a) and on the 
mean--squared displacement
for the silicon atoms (b) for the indicated temperatures.
}
\end{figure}

It is also remarkable that the evolution of 
$D_{\rm Y}$ and $\Delta$ with the cutoff is not completely trivial, 
but exhibits some structure. This is best seen 
in the data of Fig.~\ref{fig3}b and c at $T=3580$~K, for 
which we have taken more data points.
This behavior can be understood from the pair correlation function for the
SiSi correlations (Fig.~\ref{fig6}a). The comparison of Fig.~\ref{fig6}a
to Figs.~\ref{fig3}b and c,  shows that the locations of shoulders
or even maxima in $D_{\rm Y}$ and $\Delta$ are at the locations of
the minima in $g_{\rm SiSi}(r)$. This makes sense because the minima
in $g_{\rm SiSi}(r)$ correspond to distances of SiSi pairs that are
relatively unlikely and so around these distances the change in the
effective screening is weaker than for separations of SiSi pairs at
which strong structural correlations occur.

The pair correlation function $g_{\rm SiSi}(r)$ in Fig.~\ref{fig6}b
exhibits a similar behavior as in the case of the Wolf method,
although the convergence towards the Ewald
result is slower than for the Wolf method. For interparticle distances
$r>4$\,\AA~relatively strong deviations from the Ewald result can be
observed for $r_c=6.0$\,\AA. Note that the first peak in $g_{\rm SiSi}(r)$
which is not shown in Fig.~\ref{fig6}b is also well reproduced for
$r_c=6.0$\,\AA. For $r_c=10.17$\,\AA~the agreement with the Ewald result
extends to the second peak in $g_{\rm SiSi}(r)$, while for interparticle
distances $r>7$\,\AA~the Yukawa result is out--of--phase with respect
to the Ewald result. 

The dynamic properties of the Yukawa potentials are
represented in Fig.~\ref{fig6}b by the mean squared displacement $\langle
r_{\alpha}^2(t) \rangle$ for the oxygen particles (i.e.~$\alpha={\rm O}$)
at the temperatures $T=6100$\,K and $T=3250$\,K, defined by
\begin{equation}
    \langle r_{\alpha}^2(t) \rangle  = \left\langle 
           \frac{1}{N_\alpha} \sum_{i=1}^{N_\alpha}
            |{\bf r}_i(t) - {\bf r}_i(0) |^2 \right\rangle.
\end{equation}
We find similar results for Si atoms.
For different values of $r_c$ a similar qualitative behavior at the high
and the low temperature is seen, albeit effects are more pronounced at the low
temperature. In the diffusive regime, the dynamics becomes faster when
decreasing the cutoff $r_c$. Thus, as in the case of the Wolf method, 
a change in long--range structural correlations is accompanied by a faster
diffusion of the particles. 
In the next section, we present more extensive results 
for $r_c=10.17$\,\AA~for the Yukawa
potential. This cutoff provides a reasonable accuracy compared to
the Ewald results, although the agreement cannot be made 
as good as the Wolf case for the same cutoff value.

The simulations with the Yukawa potential (also those shown in
Figs.~\ref{fig6}) were done for systems of 1152 particles at a total
mass density of 2.37\,g/cm$^3$, i.e.~the same density that we also used
for the simulations in which the Ewald sums and those in which the Wolf
method was applied. Furthermore the results presented in the next section 
have been obtained considering  average values for the Yukawa parameters, i.e.
$D_{\rm Y}=1.07$ and $\Delta=5.649$~\AA.

\section{Detailed comparison to the long--range model}
\label{results}

In this section, we present a detailed comparison of the static and
dynamic behavior of the Wolf and Yukawa finite--range potentials for
liquid silica and the one obtained with Ewald summation using 
parameters described in the previous section.

\subsection{Static properties}

\begin{figure}
\psfig{file=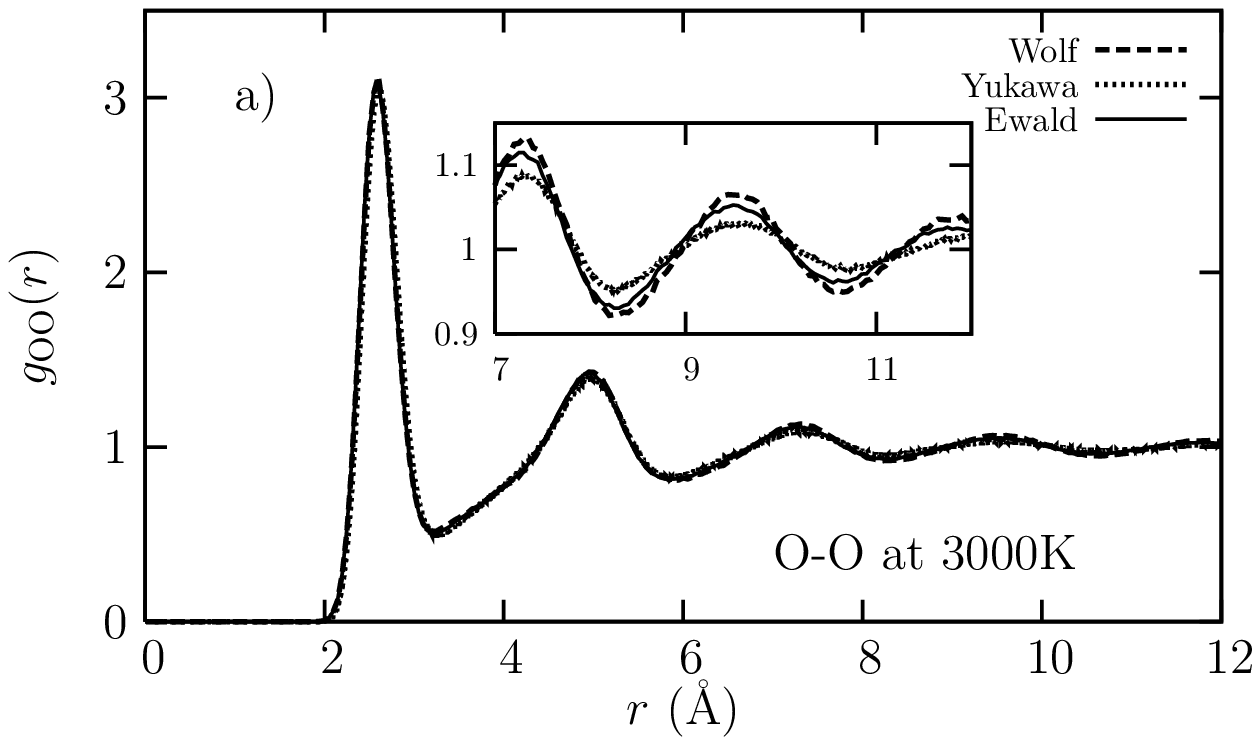,width=8.5cm}
\psfig{file=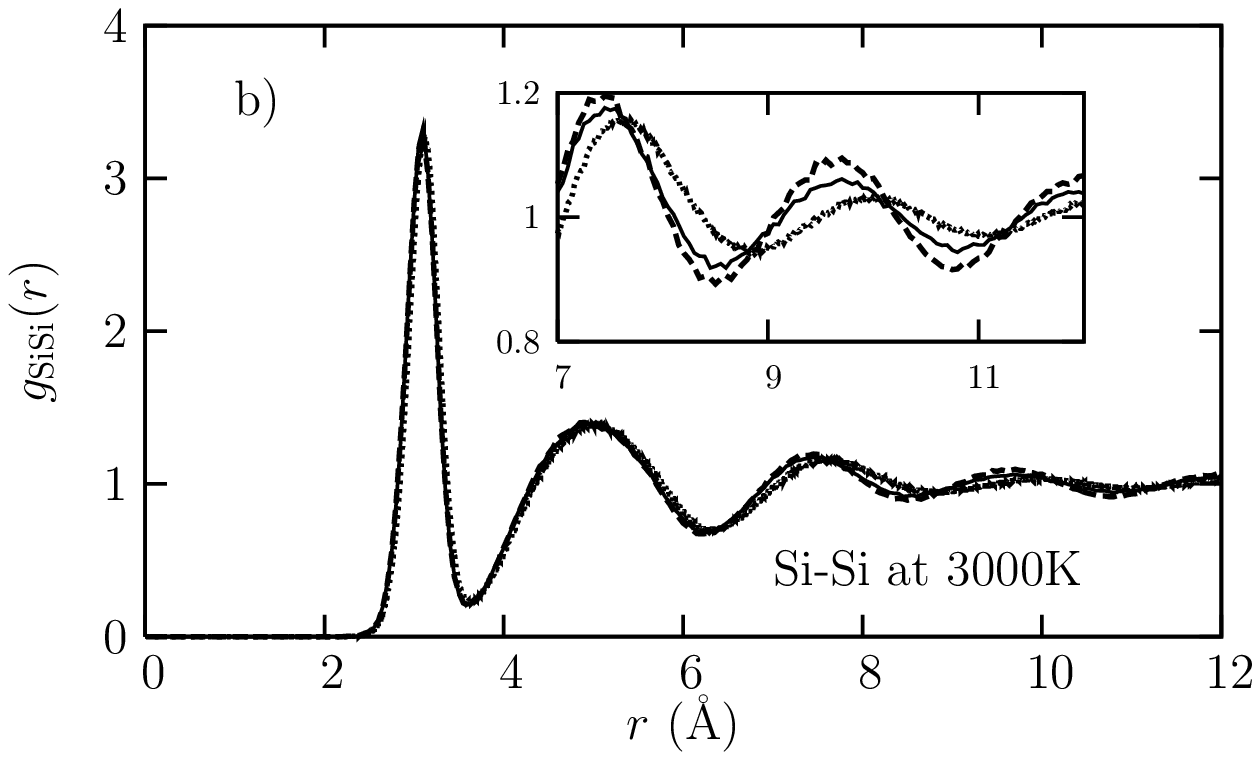,width=8.5cm}
\psfig{file=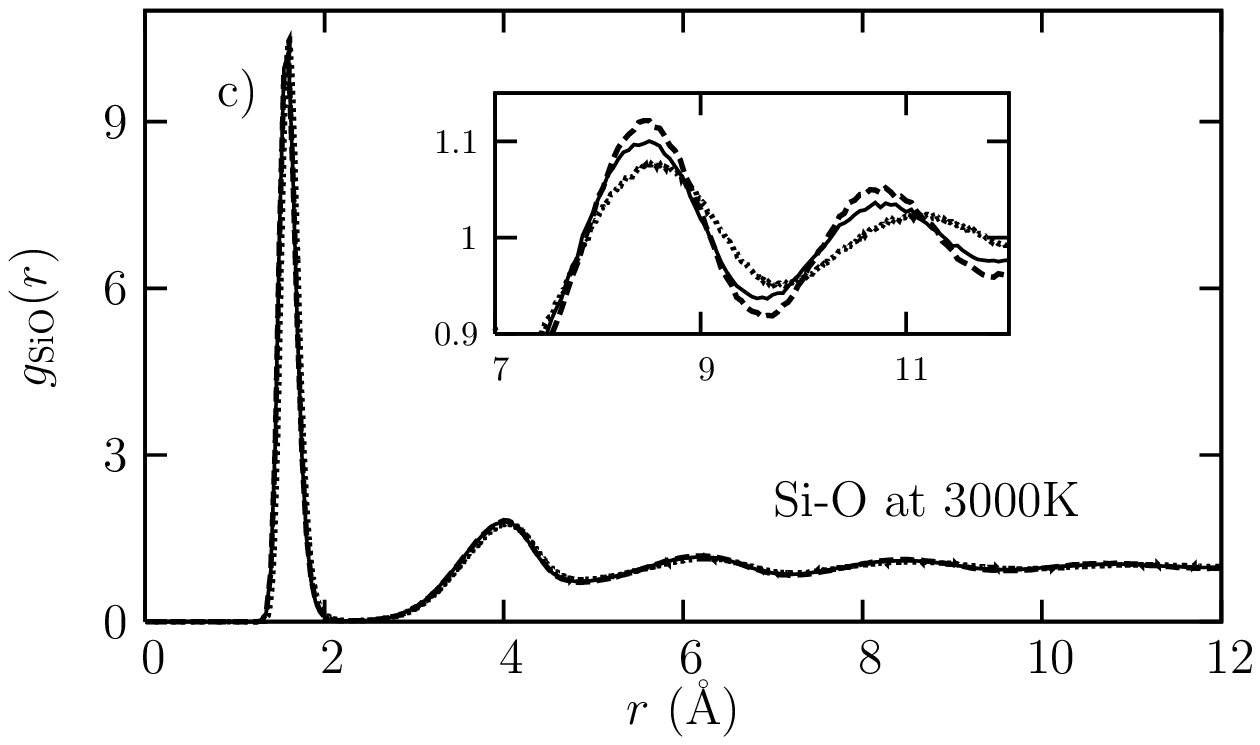,width=8.5cm}
\caption{\label{gr} Static pair correlation functions for 
Ewald, Wolf and Yukawa simulations for $T=3000$~K ($r_c=10.17$~\AA). 
The overall structure of the liquid is the same for the three 
simulations, although small deviations are observed 
at large distances, as shown in the insets.}
\end{figure}
In Fig.~\ref{gr} we show the static pair correlation functions for Si--Si,
O--O and Si--O pairs at a single, low temperature, $T=3000$~K. At first
sight, it is obvious that there is a fairly good agreement between the
three sets of data. Looking at the data more closely, one can see that
although short--range structure, $0 < r < 5$\,\AA, is well reproduced
by the two finite--range potentials, small deviations are present
at larger distances, $r> 7$\,\AA, as evidenced in the three insets
of Fig.~\ref{gr}. It is interesting that Wolf and Yukawa potentials
produce deviations with opposite trends.  Wolf data reveal a liquid which
is more structured at large distances than in the Ewald simulations,
while the Yukawa potential produces less structured pair correlation
functions.  
We do not have a simple physical explanation for this
opposite tendency. We also remark that large distance oscillations in
the Yukawa potential are slightly out--of--phase with the Ewald results,
as is best seen for the Si--Si correlations (see also Fig.~\ref{fig6}b).
But overall we conclude that the local structure of silica is very well
reproduced by the finite--range potentials. Large distance correlations
are more sensitive to the truncation of the Coulombic interaction, but
reasonable agreement can nevertheless be obtained for the parameters
chosen in Sec.~\ref{models}.

\begin{figure}
\psfig{file=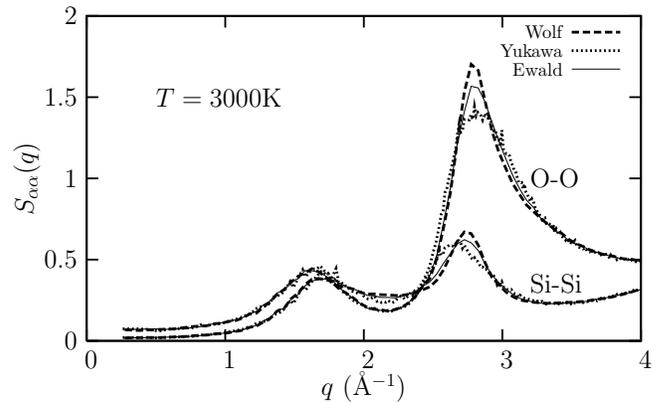,width=8.5cm}
\caption{\label{gr2}  Partial static structure 
factor for O--O and Si--Si correlations
at $T=3000$\,K, and for the three simulations methods. The overall structure
of the fluid is correct, with small deviations close to the peak 
at $2.9$\,\AA$^{-1}$.}
\end{figure}

\begin{figure}
\psfig{file=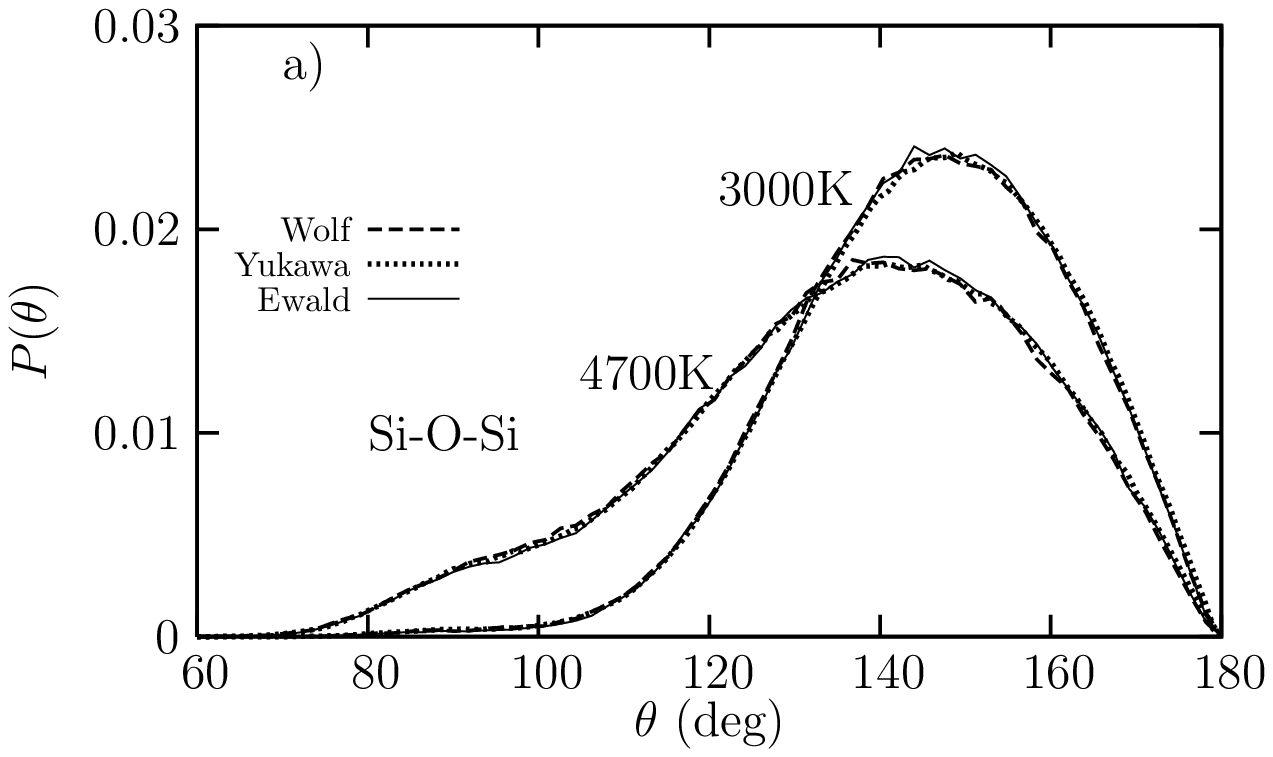,width=8.5cm}
\psfig{file=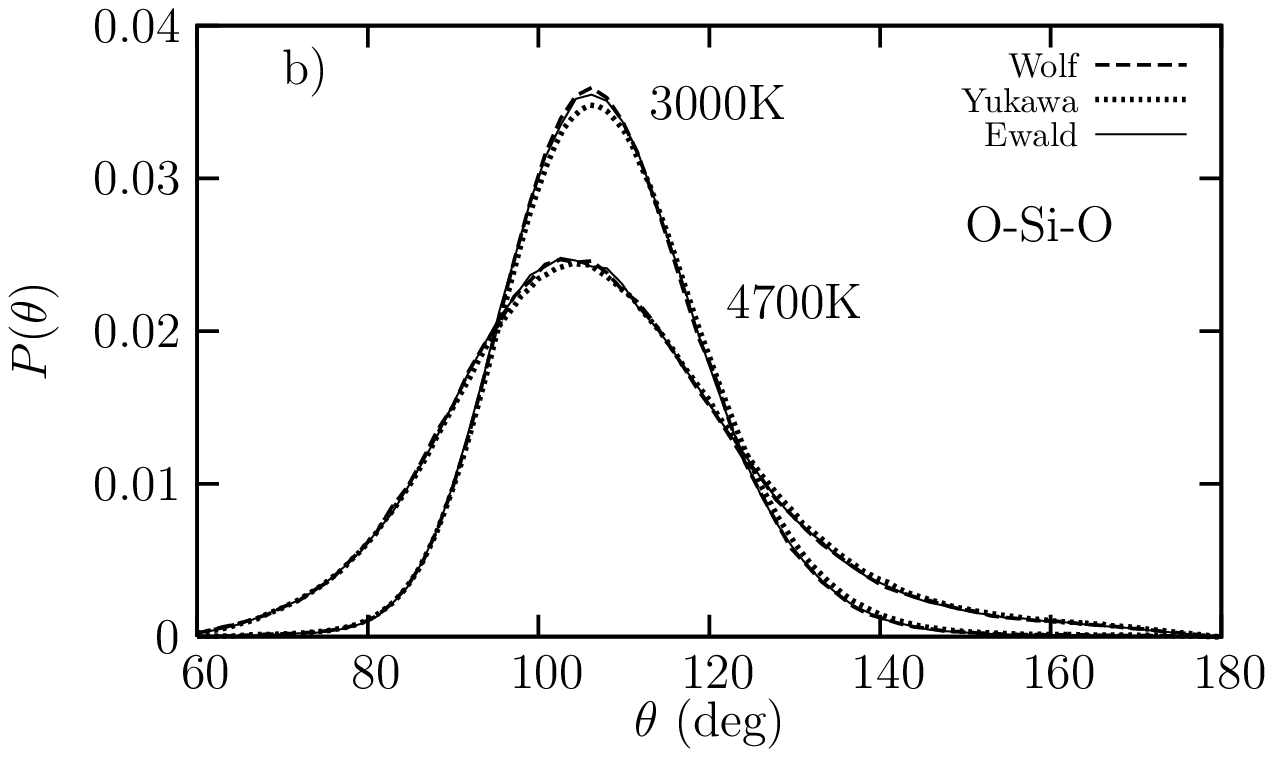,width=8.5cm}
\caption{\label{theta} Bond angle distributions 
(Si--O--Si and O--Si--O)  for 
two temperatures and for the three simulation methods.
The agreement between the three models is very good at all temperatures.}
\end{figure}

The opposite large distance behavior of the Wolf and Yukawa data are
also clearly revealed when looking at the  partial static structure factors,
as shown in Fig.~\ref{gr2}. As compared to Ewald simulations, the Wolf
data correctly reproduce $S(q)$ for both small ($q<2.5$\,\AA$^{-1}$) and
large ($q>3.5$\,\AA$^{-1}$) wavevectors, while the height of the peak at
$q\approx 2.9$\,\AA$^{-1}$ is slightly too large.  This is consistent
with the observation of a large distance structure seen in the
pair correlation functions.  Similarly, the Yukawa data underestimate
the height of this peak, while its position is also slightly incorrect
in the case of Si--Si correlations. As discussed in Sec.~\ref{models},
it is important to recall that the amplitudes of these deviations are
more pronounced at low temperatures when the fluid is more structured.

That the short--range structure of the fluid is well reproduced by our
finite--range potentials is confirmed in Fig.~\ref{theta} which presents
Si--O--Si and O--Si--O bond angle distributions for the three models.  We find
that at all temperatures the local atomic arrangements are indeed very
similar in the three cases.  
We find similar results for Si--Si--Si angle distributions.
This is consistent with the above observation
that small differences in the pair correlation functions are only seen
at very large distances.  These large distance differences do not affect
the local tetrahedral arrangement of the atoms.

\begin{figure}
\psfig{file=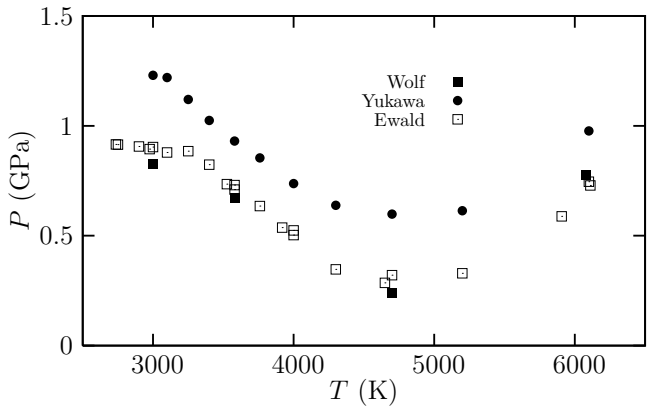,width=8.5cm}
\caption{\label{press} Temperature dependence of the pressure 
for the three potentials.}
\end{figure}

In Fig.~\ref{press} we show the temperature dependence of the pressure,
$P$. As mentioned above, we find that the pressure is very sensitive to
the truncation of the Coulombic interaction. Nevertheless, the agreement
between Wolf and Ewald results is very good, while somewhat larger
deviations are observed for the Yukawa potential. Again the sign of the
deviations is opposite for Wolf and Yukawa results.

\subsection{Dynamic properties}

\begin{figure}
\psfig{file=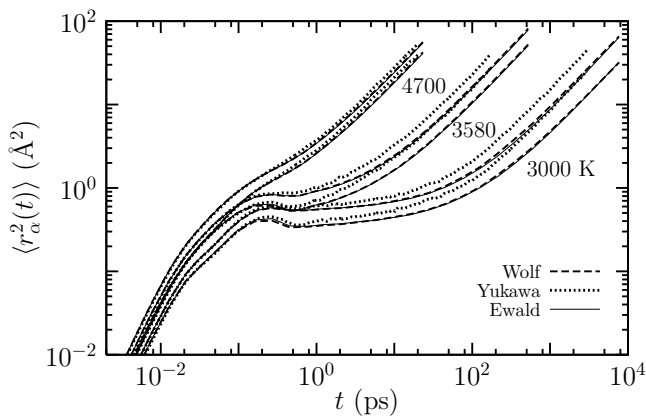,width=8.5cm}
\caption{\label{self}
Mean--squared displacements for the three simulations 
techniques at three different temperatures. At each temperature we 
show both the Si and O dynamics, the latter being the fastest. The 
agreement between Wolf and Ewald is excellent. The slowing down of 
the dynamics of the Yukawa potential is slightly less pronounced.}
\end{figure}

In Fig.~\ref{self} we compare the mean-squared displacements 
$\langle r_{\alpha}^2(t)\rangle$ for both Si and O species for several temperatures to the
Ewald results.

Within the statistical noise, we cannot observe any difference between the
Ewald and Wolf data  at the two highest temperatures. 
For $T=3000$\,K there is a small difference
between the two methods on timescales corresponding to thermal vibrations
as discussed in the context of Fig.~\ref{fig2}. A similar agreement is found
for the mean--squared displacements in Fig.~\ref{self}, although the
use of logarithmic scales somewhat obscures the little differences
between the two sets of data.  From Fig.~\ref{self} we also deduce that
the temperature evolution of the relaxation timescales and diffusion
constants is similar using both Wolf and Ewald methods.  It is striking
that despite the small differences in the structure of Ewald and Wolf
data, their dynamical behavior is in excellent agreement.
This is not true for the Yukawa potential where larger deviations 
can be noticed in Fig.~\ref{self}.

\begin{figure}
\psfig{file=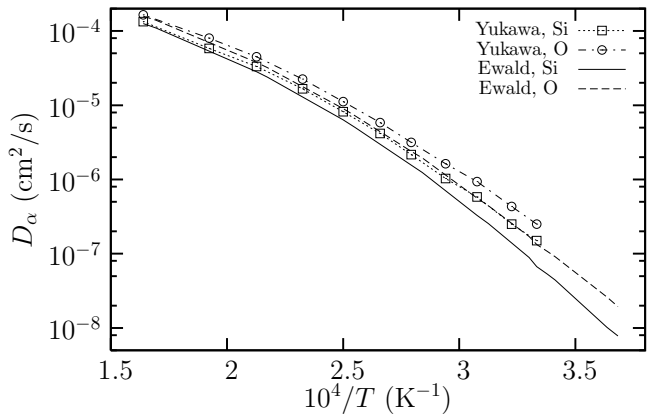,width=8.5cm}
\caption{\label{fig12} Arrhenius plots of the self--diffusion constants 
for Ewald and Yukawa methods.}
\end{figure}

The Yukawa data in Fig.~\ref{self} also imply that 
dynamic properties of the Yukawa
and the Ewald models have a slightly different temperature evolution.
This can be illustrated by the temperature dependence
of the self--diffusion constants. We have calculated the self--diffusion
constants $D_{\alpha}$ via the long--time limit of the mean squared
displacements using the Einstein relation:
\begin{equation}
  D_{\alpha} = \lim_{t \to \infty} \frac{\langle r_{\alpha}^2(t) \rangle}{6t} .
\end{equation}
Figure \ref{fig12} displays the diffusion constants in an Arrhenius
plot. Note that we do not plot the Wolf results in this figure, because
there are only minor differences between Wolf and Ewald data in the full
temperature range. Whereas at the highest temperature, $T=6100$\,K,
Yukawa and Ewald methods yield the same values for $D_{\alpha}$ within
the statistical accuracy, at lower temperatures the Yukawa values are
a factor of 2 to 3 higher than the ``exact'' Ewald values.  Note that at
low temperatures the Yukawa result for $D_{\rm Si}$ coincides with the
Ewald result for $D_{\rm O}$.  This is certainly a coincidence but it
illustrates that qualitatively the temperature dependence as obtained
from the Yukawa potential is very similar to that of the ``exact''
calculation with Ewald sums. In particular, from both methods similar
activation energies are expected for the temperature dependence of
$D_{\rm \alpha}$ in the low--temperature regime.

\begin{figure}
\psfig{file=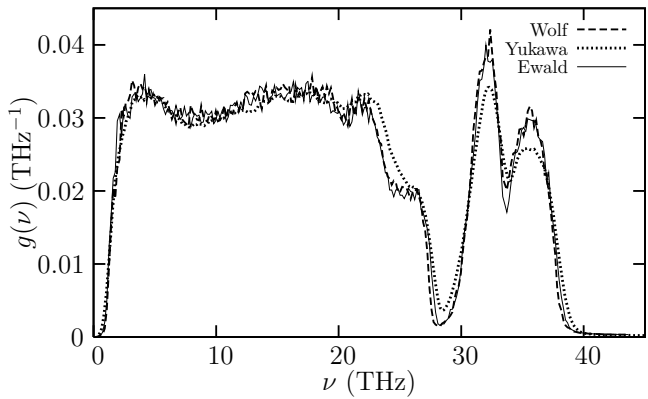,width=8.5cm}
\caption{\label{gnu} Vibrational density of states for the three potentials.}
\end{figure}

Finally we show in Fig.~\ref{gnu} that the vibrational density of states $g(\nu)$
is well reproduced by the Wolf method and, on a qualitative level,
also by the Yukawa method.  In order to compute $g(\nu)$ we quenched
10 independent liquid samples with an infinite cooling rate, i.e. a steepest
descent, down to
$T=300$\,K followed by an annealing during 80\,ps. We then measured
the velocity autocorrelation function and obtained $g(\nu)$ from its
Fourier transform \cite{dove}.  The liquid starting configurations
were well--equilibrated samples at $T=2715$\,K for the Ewald as well
as the Wolf case and at $T=3000$\,K for the Yukawa case.  As we see in
Fig.~\ref{gnu}, Ewald and Wolf results for $g(\nu)$ are in very good
agreement, while the Yukawa method yields to slight deviations from the
Ewald calculations, in particular at high frequencies ($\nu>30$\,THz)
where $g(\nu)$ exhibits a double--peak structure. The latter peaks
are due to inter--tetrahedral stretching modes. In the past it has been
shown that cooling rate effects affect especially the amplitude of
the high--frequency peaks \cite{vollmayr96,horbach99_2}. Thus, part of
the deviations seen in the Yukawa results might be due to the different
cooling history that we used for the Yukawa case. We also mention that in
the frequency range 5\,THz$\le \nu \le$25\,THz the vibrational dynamics in
the BKS model is quite unrealistic, as has been revealed by the comparison
to Car--Parrinello MD simulations \cite{benoit02}. The results of the
present study suggest that an improvement of the BKS model with respect to
vibrational properties can be also achieved by a finite--range potential.

\section{Conclusion}
\label{conclusion}

Extensive MD simulations have been used to study finite--range
approximations of the BKS model for silica. We have demonstrated
that both the Wolf method and a screened Coulomb (Yukawa) potential 
can be used to approximate the Coulomb interactions in BKS silica.
In order to obtain quantitative agreement with Ewald sums,
a cutoff of about 10\,\AA~has to be used for the Wolf potential.
Such a potential cannot be classified as a short--range potential
since its spatial extent corresponds to about three connected
SiO$_4$ tetrahedra. However, due to its finite range,
the Wolf potential can be efficiently applied to problems where
Ewald sums become inefficient, such as quasi--two--dimensional geometries,
MC simulations~\cite{berthier07,berthier07b}, 
or MD simulations with a very large number of particles.  
In the case of the Yukawa potential larger cutoffs have to be used
to yield quantitative agreement with Ewald results. However, we have
to keep in mind that we have simply reparametrized the Coulomb interaction
term of the BKS model. An interesting possibility would be to 
perform a more complete reparametrization of the BKS potential, 
so that a smaller cutoff value could be used, making simulations 
even more efficient.

\begin{acknowledgments}

We are grateful to K.~Binder and D. Reichman 
for stimulating discussions.  We gratefully
acknowledge financial support by Schott Glas and computing time on the
JUMP at the NIC J\"ulich, and on IBM/SP at CINES in Montpellier.
LB is partially funded by the Joint Theory Institute at the Argonne
National Laboratory and the University of Chicago.
 
\end{acknowledgments}

\end{document}